\documentclass[a4paper,11pt,preprint,showpacs,superscriptaddress,preprintnumbers,prd]{revtex4}
\usepackage{graphicx,amsmath,dcolumn,bm}
\usepackage{hyperref}
\usepackage{epstopdf}
\begin{document}
\title{\Large{\bf{Bjorken variable and scale dependence of quark transport coefficient in semi-inclusive lepton-production of hadron off nuclei}}}

\author{Tian-Xing Bai}
\email[E-mail:]{txbai@hebtu.edu.cn}
\affiliation{College of Physics and Hebei Advanced Thin Film Laboratory, Hebei Normal University, Shijiazhuang 050024, P.R.China}

\author{Chun-Gui Duan}
\email[E-mail:]{duancg@hebtu.edu.cn}
\affiliation{College of Physics and Hebei Advanced Thin Film Laboratory, Hebei Normal University, Shijiazhuang 050024, P.R.China}

\begin{abstract}

Nuclear modification of hadron production in deep inelastic lepton-nucleus scattering can be applied to study the parton propagation mechanism in cold nuclear matter. By means of the analytic parameterization of quenching weight based on BDMPS formalism with the target nuclear geometry effect, the leading-order computations for hadron multiplicity ratios are performed with comparison to the HERMES charged pions production data on the quarks hadronization occurring outside the nucleus.
The relation is discovered between quark transport
coefficient and the measurable kinematic variables in deep inelastic scattering.
Four models are proposed on the quark transport coefficient.
The constant model, the power-law model and the double power-law model can be ruled out because of the experimental fact
that the transverse momentum broadening increases as a function of the photon virtuality $Q^{2}$. The quark transport coefficient is determined as a function of the Bjorken variable $x$ and scale $Q^2$.  The trend of quark transport coefficient in respect of Bjorken variable $x$ and scale $Q^2$ is qualitatively in partial agreement with HERMES experimental data on transverse momentum broadening. It is hoped that our effort is conducive to understanding of jet quenching phenomenon in relativistic heavy ion collisions.

\vskip 0.1cm



\end{abstract}

\maketitle

\newpage
\vskip 0.5cm

\section{Introduction}

Nuclear modification of hadron production in deep inelastic lepton-nucleus scattering[1-9] can be used to study the parton propagation mechanism in cold nuclear matter. For lepton-nucleus deep inelastic scattering, the interaction point is well defined between the virtual photon and a bounded quark in a nucleus, since the virtual photon energy is measured by the difference of the incoming and outgoing lepton energies.
In addition, the properties of target nuclei in which the struck quark propagates, i.e. cold nuclear matter, is well known. Therefore, the precise knowledge of parton propagation mechanism obtained from the semi-inclusive lepton-production of hadron off nuclei under "clean" conditions, is essential to explore the similar process appearing in relativistic heavy ion collisions. The mechanisms by which the traversing partons lose energy to, and interact with the quark-gluon plasma, are still not fully understood[10-13].

According to the picture of multiple parton scattering by Baier, et al.[14-15](BDMPS hereafter), the energy loss for an energetic parton propagating in a nucleus is dominated by induced gluon bremsstrahlung. The parton energy loss is determined by the so-called "transport coefficient"
\begin{equation}
\hat{q}_{R}= \rho \int dq^{2}q^{2}\frac{d\sigma_{R}}{dq^{2}},
\end{equation}
where $\rho$ is the nuclear matter density, $q^{2}$ is the two-dimensional momentum transfer squared in a propagating parton-nucleon scattering, and $R$ is the color representation of the propagating parton in SU(3). By means of the transport coefficient, the parton transverse momentum broadening due to medium induced radiation reads[16]
\begin{equation}
\left\langle q^{2}_{\perp}\right\rangle=\hat{q}_{R}L ,
\end{equation}
with the length $L$ crossed by the propagating parton in the target. The transport coefficient in the target nucleus can be calculated in terms of the gluon distribution function in a nucleon[14],
\begin{equation}
\hat{q}_{R}=\frac{4\pi^{2}\alpha_{\rm s}(Q^{2}_{\rm G})C_{R}}{N^{2}_{c}-1}\rho x_{\rm G}G(x_{\rm G},Q^{2}_{\rm G}),
\end{equation}
where $C_{R}$ is the corresponding colour factor for quark or gluon, the number of colors $N_{c}=3$, the virtuality $Q^{2}_{\rm G}=\hat{q}_{R}L$. The strong coupling constant $\alpha_{\rm s}$ and the gluon distribution function $G(x_{\rm G},\hat{q}_{R}L)$ depend on the scale $\hat{q}_{R}L$. The theoretical calculation indicates that Bjorken variable $x_{\rm G}\ll1$ in the gluon distribution.

In 2010, the HERMES collaboration[17] reported firstly their detailed determination of the transverse momentum broadening of hadrons in semi-inclusive deep inelastic scattering for charged pions and positively-charged kaons on He, Ne, Kr, and Xe target. The average squared transverse momentum is clearly observed to increase with atomic mass number $A$. The broadening increases as a function of the photon virtuality $Q^{2}$ and Bjorken scaling variable $x$, and remains constant as a function of both the virtual photon energy $\nu$ and the fractional hadron energy $z$, except that it vanishes as $z$ approaches unity. The experimental data shows that $\hat{q}L<1$ GeV$^{2}$. Hereafter we shall label $\hat{q}$ the quark transport coefficient.

As for the quark transport coefficient $\hat{q}$, although Bjorken variable $x_{\rm G}$ is estimated in theory[14], $x_{\rm G}$ in the definition (3) is an unobservable kinematic variable in experiment. The unmeasurable $x_{\rm G}$ makes $x_{\rm G}G(x_{\rm G},\hat{q}_{R}L)$ unknown. The unknown $x_{\rm G}G(x_{\rm G},\hat{q}_{R}L)$ prevents us from making certain the characteristic of quark transport coefficient $\hat{q}$. This often limits the predictive power of theory. Therefore, in order to realize the details about quark transport coefficient, one would like to know whether the unmeasurable $x_{\rm G}$ can be replaced by the measurable kinematic variable in deep inelastic scattering. The purpose of this paper is to establish the relation between quark transport coefficient $\hat{q}$ and the measurable kinematic variable in deep inelastic scattering.

In fact, as early as in 2014, the scale dependent of jet transport parameter has been derived theoretically[18].
Following the similar idea, by means of the experimental data on transverse momentum broadening from semi-inclusive electron-nucleus deep inelastic scattering, Drell-Yan dilepton and heavy quarkonium production in proton-nucleus collisions, as well as the nuclear structure functions in deep inelastic scattering, the kinematic and scale dependence of jet transport coefficient in cold nuclear matter was determined in the framework of the generalized QCD factorization formalism[19]. By fitting the world data on nuclear transverse momentum broadening from Drell-Yan and quarkonium production for pion and proton incident on nucleus, the transport coefficient parameterization as a function of Bjorken variable was extracted[20]. In order to research the underlying structure of the quark-gluon plasma (QGP), the jet transport coefficient was suggested to be expressed in terms of a parton distribution function with energy and scale dependence[21,22].
Recently, by using the nuclear parton distributions determined only with lepton-nuclear
deep inelastic scattering experimental data and the analytic parameterization of quenching
weight based on BDMPS formalism[23,24], a phenomenological analysis of the nuclear Drell-Yan
differential cross section ratio as a function of Feynman variable was performed from Fermilab
E906[25] and E866[26] experimental data.
The relation was discovered
between quark transport coefficient and the measurable kinematic variables in Drell-Yan reaction.
The Bjorken variable and scale dependence of quark
transport coefficient was explored[27].

In our preceding articles[28-30], we have researched the nuclear modifications of hadron production in semi-inclusive deep inelastic scattering in a parton energy loss model with the form of the mean energy loss. Our theoretical results show that the nuclear effects on parton distribution functions can be neglected. It is found that the theoretical results considering the nuclear modification of fragmentation functions due to quark energy loss are in good agreement with the selected experimental data. In the recent work[31], we employed the two-dimensional data from the HERMES Collaboration on the multiplicity ratios for positively charged pions production on neon nucleus with respect to deuterium target[9]. The leading-order hadron multiplicity ratios were calculated by means of Salgado-Wiedemann(SW) quenching weights[32] and the analytic parameterizations of quenching weights[33] based on BDMPS formalism. We studied the target nuclear geometry effect on hadron production, and compared our calculated results with the HERMES positively charged pions production data with the quarks hadronization occurring outside the nucleus. The extracted transport parameters from the global fit are shown to be $\hat{q}=0.74\pm0.03$ GeV$^2$/fm for the SW quenching weight without the finite energy corrections, and $\hat{q}=0.20\pm0.02$ GeV$^2$/fm for the analytic parameterization of BDMPS quenching weight without the quark energy $E$ dependence, respectively. We found that the nuclear geometry effect has a significant impact on the quark transport coefficient in cold nuclear matter.

In this paper, the experimental data with the quarks hadronization occurring outside the target nucleus, are selected on the multiplicity ratios from HERMES for charged pions production between neon and deuteron targets[9]. The BDMPS quenching weight[14,33] will be employed with considering the quark energy loss and nuclear geometry effect. The Bjorken variable and scale dependence of quark transport coefficient is investigated in semi-inclusive lepton-production of hadron off nuclei. It is hoped to gain new knowledge about the characteristic of the quark transport coefficient $\hat{q}$ in a cold nuclear medium.

The paper is organized as follows. In Section II we give details of the formalism for the hadron multiplicity in semi-inclusive lepton-production of hadron off the nucleus and nuclear modification of the fragmentation functions owing to quark energy loss and nuclear geometry effect. The results and discussion obtained are presented in Section III. Finally, we summarize our findings in Section IV.

\section{The hadron multiplicity in semi-inclusive deep inelastic scattering on the nucleus}

According to the factorization theorem[34] and using leading order QCD approximation, we can write the hadron multiplicity in the semi-inclusive deep inelastic scattering off a nucleus,
\begin{equation}
\frac{1}{N^{\rm DIS}_{A}}\frac{dN^{h}_{A}}{dzd\nu}=\frac{1}{\sigma^{\rm lA}}\int dx\sum_{f}e^{2}_{f}q^{A}_{f}(x,Q^{2})\frac{d\sigma^{\rm lq}}{dxd\nu}D^{A}_{f|h}(z,Q^{2}),
\end{equation}
\begin{equation}
\sigma^{\rm lA}=\int dxd\nu\sum_{f}e^{2}_{f}q^{A}_{f}(x,Q^{2})\frac{d\sigma^{\rm lq}}{dxd\nu}.
\end{equation}
Here, $N_{A}^{h}$ ($N_{A}^{\rm DIS}$) is the yield of semi-inclusive(inclusive) deep-inelastic scattering leptons on nucleus $A$, $\sigma^{\rm lA}$ is the total cross section for the lepton-nucleus deep inelastic scattering, $e_f$ is the electric charge of the quark with flavor $f$, $q^{A}_{f}(x,Q^{2})$ is the nuclear quark distribution function, $D^{A}_{f| h}(z,Q^{2})$ is the nuclear modified fragmentation function of a quark of flavour $f$ into a hadron $h$. In above two equations, the differential cross section for lepton-quark scattering at leading order,
\begin{equation}
\frac{d\sigma^{\rm lq}}{dxd\nu}=M_{\rm N}x\frac{4\pi\alpha_{\rm em}^{2}}{Q^{4}}[1+(1-y)^{2}],
\end{equation}
where $M_{\rm N}$ and $\alpha_{\rm em}$ are respectively the nucleon mass and the electromagnetic coupling constant, $y$ is the fraction of energy of the incoming lepton which is transferred to the virtual photon.

In semi-inclusive deep inelastic scattering, the energetic parton propagating in a target nucleus suffers multiple scattering and radiate soft gluons. At the time of hadronization, the quark energy is reduced from $E=\nu$ to $E'=\nu-\Delta E$ in which $\Delta E$ the quark energy loss in the nuclear medium. The fragmentation function in the nuclear medium can be expressed as[35],
\begin{equation}
D^{A}_{f|h}(z,Q^{2})=\int_{0}^{(1-z)\nu}d(\Delta E) P(\Delta E,\omega_{c},L)\frac{1}{1-\Delta E/\nu}D_{f|h}(z',Q^{2}),
\end{equation}
with $z'=z/(1-\Delta E/\nu)$. $D_{f|h}(z',Q^{2})$ is the standard(vacuum) fragmentation function of a quark of flavour $f$ into a hadron $h$. $P(\Delta E,\omega_{c},L)$, the so-called quenching weight, is the probability that the radiated gluons carry altogether a given energy $\Delta E$ for a quark with energy $E$. The characteristic gluon frequency $\omega_{c}=(1/2)\hat{q}L^{2}$ with the path length $L$ traversed by the struck quark.

For simplicity, we consider the case of the uniform hard-sphere nucleus of mass number $A$ and radius $R_A=r_0{A^{1/3}}$ with $r_0=1.12$ fm[36]. Assuming that hadronization occurs after the struck quark has escaped from target nucleus, the mean length $L$ covered by the hard quark is $L=(3/4)R_A$. If the target nuclear geometry effect is added in the semi-inclusive lepton-production of hadrons off nuclei, the colored quark produced at $y$ the coordinate along the direction of the outgoing quark, will travel the path length $L=\sqrt{R_{A}^{2}-b^{2}}-y$ with $\vec{b}$ its impact parameter[31]. Then the averaged modified fragmentation function is given by[31]
\begin{equation}
D^{A}_{f|h}(z,Q^{2})= \int d^{2}bdy\rho_{A}(\vec{b},y)\int_{0}^{(1-z)\nu}d(\Delta E)P(\Delta E,\omega_{c},L)\frac{1}{1-\Delta E/\nu}D_{f|h}(z',Q^{2}).
\end{equation}
In above equation, $\rho_A(\sqrt{b^2+y^2})=(\rho_0/A)\Theta(R_A-\sqrt{b^2+y^2})$ with $\rho_0$ is the nuclear density.

Based on the above formalism, only the transport coefficient $\hat{q}$ remains to be determined.

\section{Results and discussion}

In order to explore the characteristic of the quark transport coefficient $\hat{q}$ in a cold nuclear medium, the experimental data with quark hadronization outside the nucleus are sorted out from the two-dimensional data on the multiplicity ratio $R^{h}_{\rm M}$ for charged pions production on neon nucleus in three $z$ slices as a function of $\nu$, and in five $\nu$ slices as a function of $z$[9] under the condition $t>2R_A$ with the hadron formation time $t=z^{0.35}(1-z)\nu/\kappa$ ($\kappa=1$ GeV/fm)[7,28,37,38]. The hadron formation time is defined as the time between the moment that the quark is struck by the virtual photon and the moment that the prehadron is formed.
It is found that the number of points in selected experimental data include 8 data points for $R^{{\pi}^+}_{\rm M}(\nu)$ and $R^{{\pi}^-}_{\rm M}(\nu)$ in the $z$ region of $0.2<z<0.4$, and 30 data points for $R^{{\pi}^+}_{\rm M}(z)$ and $R^{{\pi}^-}_{\rm M}(\nu)$ in three $\nu$ regions of $14<\nu<17$ GeV, $17<\nu<20$ GeV and $20<\nu<23.5$ GeV. In total, our analysis includes 38 data points.

In calculating at leading order the hadron multiplicity ratios $R^{\pi}_{\rm M}$,
\begin{equation}
R^{\pi}_{\rm M}[\nu(z)]=\int\frac{1}{N^{\rm DIS}_{A}}\frac{dN^{h}_{A}(\nu,z)}{dzd\nu}dz(\nu)\Bigg/\int\frac{1}{N^{\rm DIS}_{\rm D}}\frac{dN^{h}_{\rm D}(\nu,z)}{dzd\nu}dz(\nu),
\end{equation}
we use the vacuum fragmentation functions given by Hirai, et al.[39], and CTEQ6L parton density in the proton[40]. As for the quenching weight, Baier, et al.(BDMPS)[23,24] have computed perturbatively the radiated gluon spectrum by hard quarks produced in QCD media. By means of the BDMPS gluon spectrum, the quenching weight $P(\Delta E,\omega_{c},L)$ can be determined[41]. The analytic parameterization[33] on BDMPS quenching weight was given,
\begin{equation}
\omega_c P(\bar{\Delta E}=\Delta E/\omega_c)=\frac{1}{\sqrt{2\,\pi}\, \sigma\,\bar{\Delta E}}\,\exp\left[-\frac{\left(\log{\bar{\Delta E}}-\mu\right)^2}{2\,\sigma^2}\right],
\end{equation}
with $\mu=-1.5$ and $\sigma=0.73$ for the outgoing quark energy $E$ independence. Moreover, the path length $L=\sqrt{R_{A}^{2}-b^{2}}-y$ in view of the target nuclear geometry effect.

In the next part of this effort, we focus on the quark transport coefficient $\hat{q}$, which is obtained by minimizing $\chi^2$. The parabolic errors of the fit parameters are given by MINUIT package[42].

If we do not consider any kinematic variable dependence of the quark transport coefficient, $\hat{q}$ is treated as a constant $\hat{q}_0$, which is below labelled the constant model. The global fit of all data gives $\hat{q}_0=0.207\pm0.017$ GeV$^2$/fm, the relative uncertainty $\delta\hat{q}_0/\hat{q}_0\simeq8\%$ and $\chi^2/{\rm ndf}=0.774$. The black solid curves in Fig.1 are our numeric results which are compared with the selected experimental data. Our calculational result shows a good fit quality.

\begin{figure}
\centering
\includegraphics*[width=15cm,height=10.5cm]{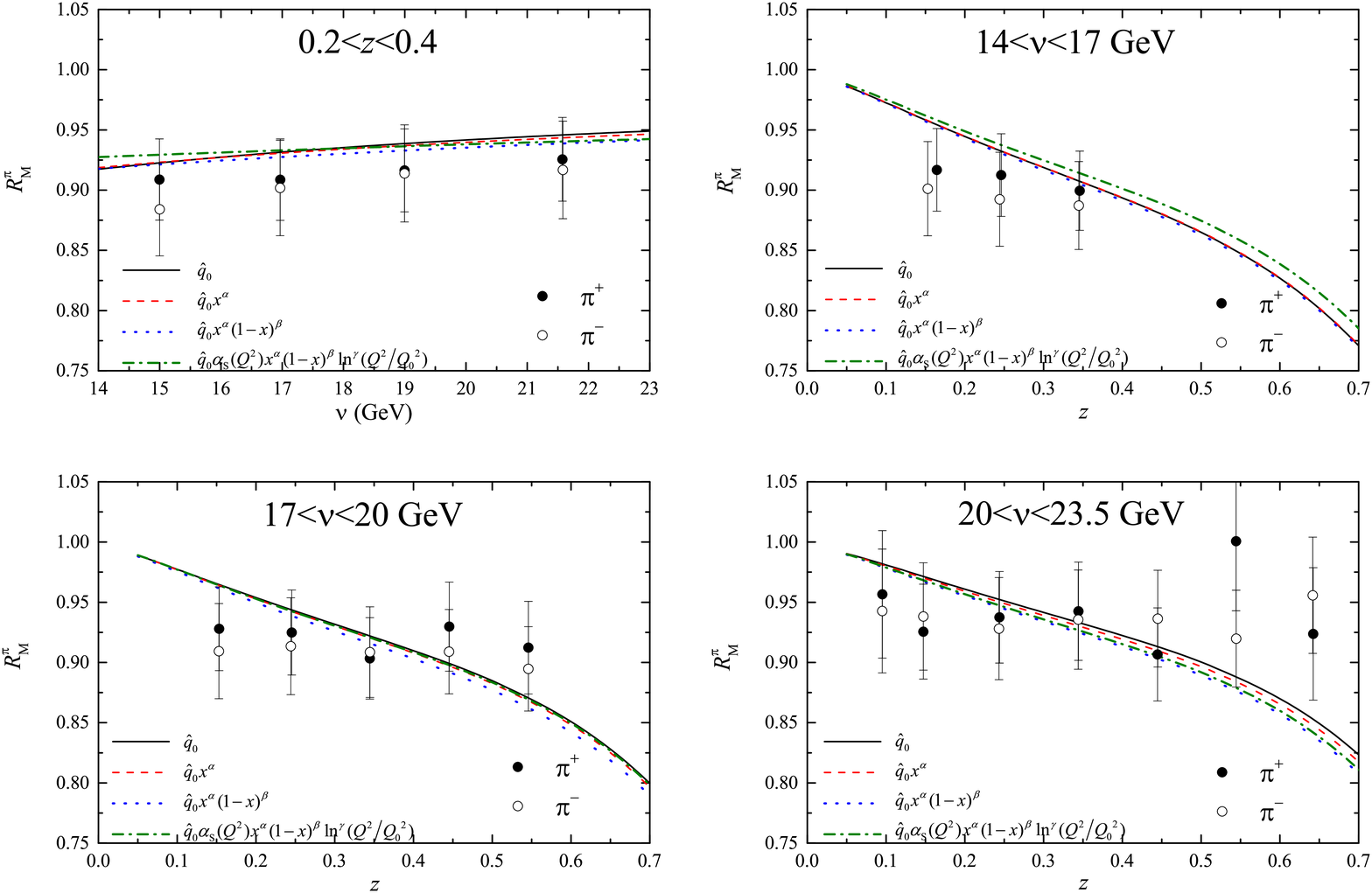}
\vspace{-0.1cm}
\caption{The calculated multiplicity ratios $R^{\pi}_{\rm M}$ as a function of $\nu$ in the region of $0.2<z<0.4$ and as a function of $z$ in $14<\nu<17$ GeV, $17<\nu<20$ GeV and $20<\nu<23.5$ GeV for charged pions production on neon target from the parameterization of BDMPS quenching weigh for the quark energy $E$ independence. The different lines correspond respectively to the results from $\hat{q}$ in four different models. The HERMES data[9] are shown with the total uncertainty (statistical plus systematic, added quadratically).}
\end{figure}

Although the quark transport coefficient $\hat{q}$ as a constant gives a high quality fit, the detail of the quark transport coefficient $\hat{q}$ is still unclear. From the formula (3) for the quark transport coefficient, $\hat{q}$ depends on the gluon distribution function $G(x_{\rm G},Q^2_{\rm G})$, which is a function of the Bjorken variable $x_{\rm G}$ and the virtuality $Q^2_{\rm G}$. Following Ref.[14], a fast parton of momentum $p$ traversing cold nuclear matter interacts with a target nucleon of momentum $P$ by the exchanged gluon of momentum $q$. In the target nucleus rest frame, using light-cone variables, the Bjorken variable $x_{\rm G}$ in the gluon distribution is given by
\begin{equation}
x_{\rm G}\equiv\frac{q^{+}}{P^{+}}=\frac{q^{+}}{M_{\rm N}}.
\end{equation}
In the semi-inclusive deep inelastic lepton-nucleus scattering, the Bjorken scaling variable $x$ is defined as
\begin{equation}
x\equiv\frac{Q^{2}}{2M_{\rm N}\nu}.
\end{equation}
It is important to remember that $x_{\rm G}$ is an unobservable quantity, $x$ is a kinematic variable measured in the deep inelastic scattering experiment. Thus, according to the above two definitions,
\begin{equation}
x_{\rm G}=\frac{2\nu q^{+}}{Q^{2}}x.
\end{equation}
In this way, the unmeasurable $x_{\rm G}$ can be characterized by the measurable Bjorken variable $x$.

As mentioned in Section I, $Q^{2}_{\rm G}<1$ GeV$^2$ in the gluon distribution from the quark transport coefficient. When the photon wavelength $1/Q_{\rm G}$ reaches the size of the nucleon with radius $R_{\rm N}$, the so-called saturation occurs. At saturation scale $Q_{\rm S}$, the parton system becomes to look dense to the probe. In this regime, the number of gluons is given by[43]
\begin{equation}
x_{\rm G} G(x_{\rm G},Q^2_{\rm S})\sim\frac{\pi}{\alpha_{\rm s}(Q^2_{\rm S})}Q^2_{\rm S}R^2_{\rm N}.
\end{equation}
The saturation scale[44,45]
\begin{equation}
Q^2_{\rm S}(x_{\rm G})\sim x^{-\lambda}_{\rm G}.
\end{equation}
Therefore, we can thus extract Bjorken variable $x$ dependence of the quark transport coefficient by writing
\begin{equation}
\hat{q}=\hat{q_0}x^{\alpha},
\end{equation}
which is referred to as the power-law model hereafter. Two parameters of the fits with the power-law model are: $\hat{q_0}$ and $\alpha$.

In the power-law model, the global fit to the selected experimental data shows that $\hat{q}_0=0.101\pm0.003$ GeV$^2$/fm with the relative uncertainty $\delta\hat{q}_0/\hat{q}_0\simeq3\%$, $\alpha=-0.288\pm0.033$ with $\delta\alpha/\alpha\simeq11\%$, and $\chi^2/{\rm ndf}=0.825$. As can be seen from Fig.1, our theoretical results(red dashed lines) are in good agreement with the experimental data. From Fig.2, we can find that the quark transport coefficient $\hat{q}$ decreases with the increasing Bjorken variable $x$.

\begin{figure}
\includegraphics*[width=9cm,height=7cm]{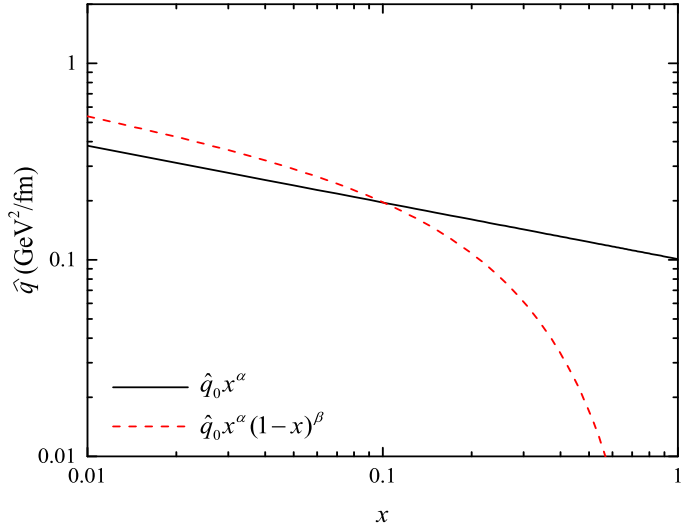}
\vspace{-0.1cm}
\caption{The extracted quark transport coefficient $\hat{q}$ as functions of Bjorken variable $x$ from the power-law model and the double power-law model, respectively.}
\end{figure}

Because of low $Q^{2}_{\rm G}$, perturbative QCD fails and non-perturbative behavior become important. The gluon distribution in the quark transport coefficient $\hat{q}$ is only an "effective" distribution. After replacing the unmeasurable $x_{\rm G}$ with the observable kinematic variable $x$ in the deep inelastic scattering experiment, and considering the intermediate and large $x$ effect, we can give Bjorken variable $x$ dependence of the quark transport coefficient,
\begin{equation}
\hat{q}=\hat{q_0}x^{\alpha}(1-x)^{\beta},
\end{equation}
which is hereafter referred to as the double power-law model. Three parameters in the double power-law model are: $\hat{q_0}$, $\alpha$ and $\beta$.

In the double power-law model, the results of our fit to the experimental data give that $\hat{q}_0=0.141\pm0.001$ GeV$^2$/fm with the relative uncertainty $\delta\hat{q}_0/\hat{q}_0\simeq1\%$, $\alpha =-0.298\pm0.033$ with $\delta\alpha/\alpha\simeq11\%$, $\beta=3.348\pm0.145$ with $\delta\beta/\beta\simeq4\%$, and $\chi^2/{\rm ndf}=0.875$. As shown in Fig.1(blue dotted lines), we see a good fit quality by taking into account the data precision. From Fig.2, we can find that the quark transport coefficient $\hat{q}$ decreases rapidly with increasing $x$ in the region $x>0.1$.

The HERMES experiment on transverse momentum broadening of hadrons produced
in semi-inclusive deep inelastic scattering on nuclei[17] reveals
that the transverse momentum broadening is a  function of Bjorken variable $x$ and photon virtuality $Q^2$.
In view of the relation (2) between transverse momentum broadening and quark transport coefficient $\hat{q}$[14,46], we can find that although the constant model, the power-law model and the double power-law model indicate a high quality fit to the selected data, these three models can not reflect the $Q^2$ dependence of quark transport coefficient $\hat{q}$. Therefore, the constant model, the power-law model and the double power-law model remain unable to give expression to complete information on quark transport coefficient $\hat{q}$.

In contrast with $x_{\rm G}$ substituted by Bjorken variable $x$, we shall conjecture that the photon virtuality $Q^2$ can replace $Q^2_{\rm G}$ in the gluon distribution though the detail is yet unknown. Thus, the strong coupling constant $\alpha_{\rm s}(Q^2_{\rm G})$ in quark transport coefficient $\hat{q}$ formula (3) can be written as $\alpha_{\rm s}(Q^2)$. After further considering the evolution of gluon distribution with the photon virtuality $Q^2$, the quark transport coefficient is assumed to be
\begin{equation}
\hat{q}=\hat{q_0}\alpha_{\rm s}(Q^2)x^{\alpha}(1-x)^{\beta}\ln^{\gamma}(Q^2/Q^2_0),
\end{equation}
which is hereafter known as the evolution model. We make $Q^2_0=1$ GeV$^2$ in order to ensure the argument in the logarithm dimensionless.
The parametrization form is  the same as that in Ref. [19].
Four parameters of the fits with the evolution model are: $\hat{q_0}$, $\alpha$, $\beta$ and $\gamma$.

\begin{table}
\caption{The parameter values of $\hat{q}$ and $\chi^{2}/{\rm ndf}$ extracted in four different models from the selected data on the hadron multiplicity ratios for charged pions produced on neon target.}
\begin{ruledtabular}
\begin{tabular*}{\hsize}
{c@{\extracolsep{0ptplus1fil}} c@{\extracolsep{0ptplus1fil}}
c@{\extracolsep{0ptplus1fil}} c@{\extracolsep{0ptplus1fil}}
c@{\extracolsep{0ptplus1fil}} c@{\extracolsep{0ptplus1fil}}}

  $\hat{q}$(GeV$^2$/fm)                                                               &$\hat{q_0}$    &$\alpha$        & $\beta$         &$\gamma$        &$\chi^{2}/{\rm ndf}$\\
  \hline
  $\hat{q_0}$                                                                         &$0.207\pm0.017$ &$              $& $             $ &$ $             &$0.774$\\
  $\hat{q_0}x^{\alpha}$                                                               &$0.101\pm0.003$ &$-0.288\pm0.033$& $             $ &$ $             &$0.825$\\
  $\hat{q_0}x^{\alpha}(1-x)^{\beta}$                                                  &$0.141\pm0.001$ &$-0.298\pm0.033$& $3.348\pm0.145$ &$ $             &$0.875$\\
  $\hat{q_0}\alpha_{\rm s}(Q^{2})x^{\alpha}(1-x)^{\beta}\ln^{\gamma}(Q^{2}/Q^{2}_0)$  &$0.401\pm0.001$ &$-0.297\pm0.022$& $3.186\pm0.055$ &$0.590\pm0.192$ &$0.967$\\

\end{tabular*}
\end{ruledtabular}
\end{table}

After performing the global fit of $\hat{q}$ to the selected experimental data. The theoretical results are compared with the selected experimental data in Fig.1. It is found that the expected multiplicity ratios shown by green dash-dotted curves are in good agreement with the experimental data. The obtained parameter values of $\hat{q}$ and $\chi^{2}/{\rm ndf}$ in our fits are summarized in Table.1. As for the evolution model, $\hat{q}_0=0.401\pm0.001$ GeV$^2$/fm with the relative uncertainty $\delta\hat{q}_0/\hat{q}_0\simeq0.2\%$, $\alpha =-0.297\pm0.022$ with $\delta\alpha/\alpha\simeq7\%$, $\beta=3.186\pm0.055$ with $\delta\beta/\beta\simeq2\%$, $\gamma=0.590\pm0.192$ with $\delta\gamma/\gamma\simeq33\%$, and $\chi^2/{\rm ndf}=0.967$. The rather small value of $\chi^{2}/{\rm ndf}$ indicates a good fit quality.

In Table.1, the determined parameter values of $\hat{q}$ and $\chi^{2}/{\rm ndf}$ are also given from the other three models. From this table and Fig.2, we can find that in the evolution model, the factor $x^{\alpha}$ reflects the small $x$ effect on the quark transport coefficient $\hat{q}$. The factor $(1-x)^{\beta}$ indicates the correction to $\hat{q}$ in the middle and large $x$ region. A narrow different of the extracted parameter $\alpha$ and $\beta$ between the evolution model and double power-law model indicates that Bjorken scaling variable $x$ part of $\hat{q}$ varying with $x$ in the evolution model follows the same trend as that from double power-law model(see the red dashed line in Fig.2).

\begin{figure}
\includegraphics*[width=15cm,height=11.5cm]{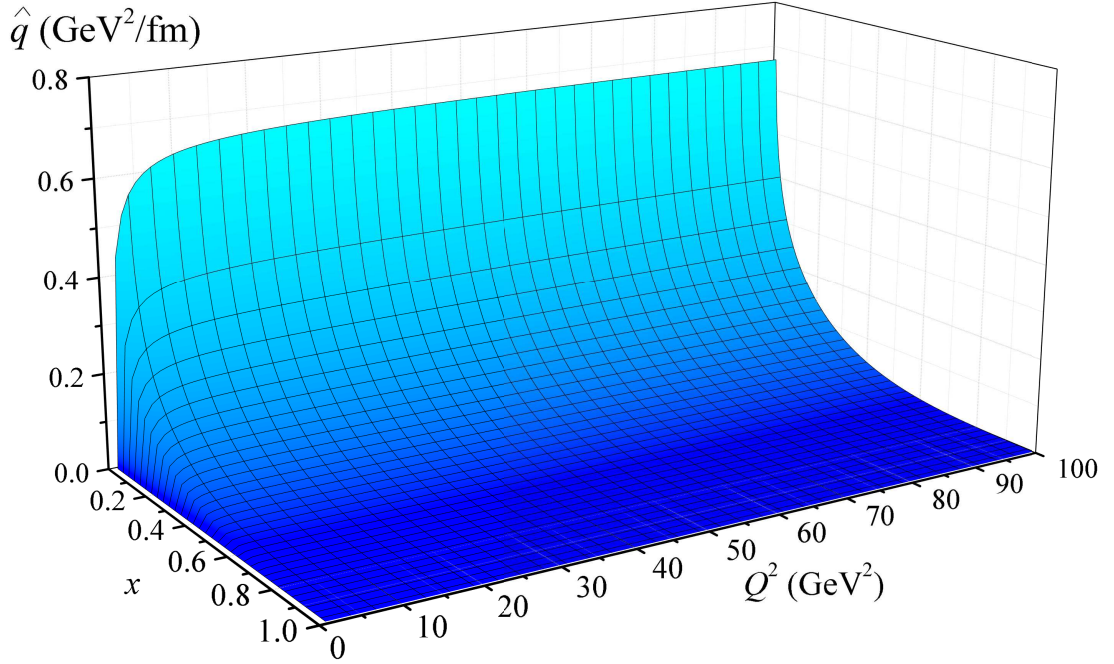}
\vspace{-0.1cm}
\caption{The obtained quark transport coefficient $\hat{q}$ as a function of Bjorken variable $x$ and scale $Q^2$.}
\end{figure}

\begin{figure}
\includegraphics*[width=9cm,height=7cm]{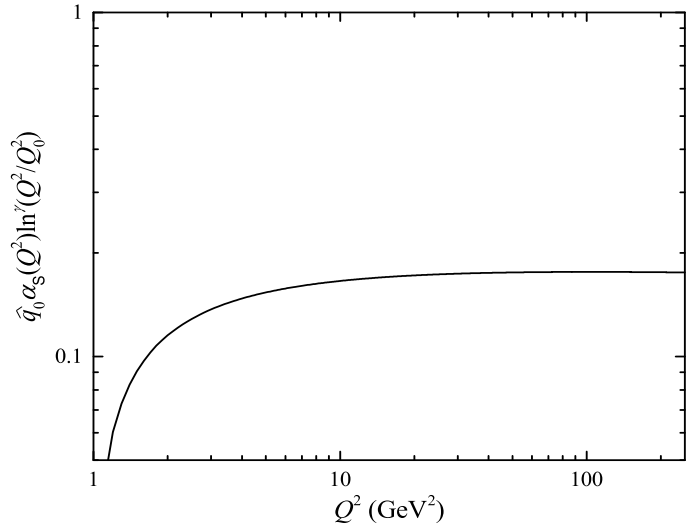}
\vspace{-0.1cm}
\caption{The part $\hat{q}_0\alpha_{\rm s}\ln^\gamma(Q^2/Q_0^2)$ in $\hat{q}$ varying with scale $Q^2$ in the evolution model.}
\end{figure}

In Fig.3, Bjorken variable $x$ and photon virtuality $Q^2$ dependence is shown from the quark transport coefficient $\hat{q}$ in the evolution model. In order to investigate $\hat{q}$ dependence on the single kinematic variable, the scale $Q^2$ part of $\hat{q}$ varying with $Q^2$ is given in Fig.4. It is found that the factor $\hat{q_0}\alpha_{\rm s}(Q^2)\ln^{\gamma}(Q^2/Q^2_0)$ increases with $Q^2$ rapidly in the region $Q^2<6.0$ GeV$^2$, afterwards has a slow rising trend.

In Fig.5, the curves of quark transport coefficient $\hat{q}$ to Bjorken variable $x$ are drawn in four fixed photon virtuality $Q^2$ (1.4, 2.5, 3.9, and 6.5 GeV$^2$). Four $(x,Q^2)$ points, (0.06, 1.4), (0.10, 2.5), (0.15, 3.9) and (0.25, 6.5), correspond to the HERMES experimental measurements(see Table 2 in Ref.[17]). As shown in Fig.5, the quark transport coefficient $\hat{q}$ increases in the region $Q^2<2.5$ GeV$^2$ and $x<0.1$ as a function of $ Q^2$ and $x$. Therefore, the trend of $\hat{q}$ in respect of the Bjorken variable $x$ and scale $Q^2$ is qualitatively in partial agreement with HERMES experimental data on transverse momentum broadening[17].

\begin{figure}
\includegraphics*[width=9cm,height=5.7cm]{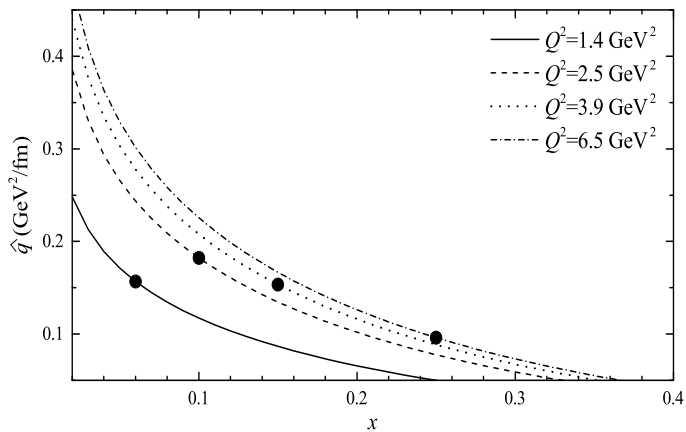}
\vspace{-0.1cm}
\caption{The quark transport coefficient $\hat{q}$ to Bjorken variable $x$ in four fixed $Q^2$ with four $(x,Q^2)$ points from the HERMES experimental measurements.}
\end{figure}

In our present effort, the used experimental data on the hadron multiplicity ratios $R^{\pi}_{\rm M}$ are from HERMES Collaboration[9] by using 27.6 GeV positron and electron beams stored in HERA at DESY. The experimental data were acquired by the following kinematic conditions: $x=0.02-1.0$ and $Q^2=1-45$ GeV$^2$. Ru, et al.[19] employed the HERMES experimental data[17] on the transverse momentum broadening for positively charged pion as a function of atomic mass number $A$, and as a function of $Q^2$, $x$ and the virtual photon energy $\nu$ from Xe target with excluding He, Ne and Kr targets. Therefore, our global fit on quark transport coefficient $\hat{q}$ has different kinematic region from those in Ref.[19]. The reason remains to be further explored on the obtained parameter values existing difference for the quark transport coefficient in the future.

\section{Summary}

The investigation into the detailed characteristic of the quark transport coefficient in a cold nuclear medium, prompts us to address the question how the quark transport coefficient depend on the measurable kinematic variables in deep inelastic scattering. We sort out the experimental data with the quark hadronization occurring outside the nucleus by the following criterion: the hadron formation time is more than twice the target nuclear radius. We take advantage of the analytic parameterization of quenching weight based on BDMPS formalism with considering the target nuclear geometry effect. Four models are provided on the quark transport coefficient $\hat{q}$. The leading order calculation of the hadron multiplicity ratio has been done, and compared with the selected HERMES experimental data. It is found that all four models display high quality fit.
However, the constant model, the power-law model and the double power-law model can be ruled out
because of the experimental fact that the transverse momentum broadening increases a function of the photon virtuality $Q^{2}$.
The quark transport coefficient is determined as a function of Bjorken variable $x$ and scale $Q^2$.
The quark transport coefficient as a function of Bjorken variable $x$ and scale $Q^2$ demonstrates qualitatively the partial behavior in HERMES experimental data on transverse momentum broadening.

In order to gain precise knowledge about the quark transport coefficient in a cold nuclear medium, we suggest that in the follow-up experiment on the semi-inclusive hadron production in deep inelastic scattering on nuclear target, the hadron multiplicity ratios should be measured in multi-dimensional representation, which consists in a fine binning in one variable and coarser bins in other variables. Meanwhile, the future measurements on the hadron multiplicity ratio and transverse momentum broadening are needed with wider kinematic coverage (e.g. at a future Electron Ion Collider[47]).
Furthermore, on the basis of the existing researches, a better understanding of the parton propagation mechanism can be achieved in the future.
\vskip 1cm
{\bf Acknowledgments}
We thank Professor Zhi-Hui Guo for interesting and useful discussions. This work is supported in part by the National Natural Science Foundation of China(11575052, 11975090).

\end{document}